\documentclass[letterpaper,twocolumn,aps,showpacs]{revtex4}
\usepackage{epsf,epsfig,graphicx}
\usepackage{supercite}
\usepackage{subfigure}
\begin{document}
\title{Critical Temperature of Weakly Interacting Dipolar Condensates} 
\author{Konstantin Glaum$^1$,  Axel Pelster$^2$, Hagen Kleinert$^1$, and Tilman Pfau$^3$}
\affiliation{$^1$Fachbereich Physik, Freie Universit{\"a}t Berlin, 
Arnimallee 14, 14195 Berlin, Germany\\
$^2$Fachbereich Physik, Universit{\"a}t Duisburg-Essen, 
Lotharstra{{\ss}}e 1, 47048 Duisburg, Germany\\
$^3$5. Physikalisches Institut, Universit{\"a}t Stuttgart, Pfaffenwaldring 
57, 70569 Stuttgart, Germany}
\date{\today}
\begin{abstract}
We calculate perturbatively the effect of a dipolar interaction upon 
the Bose-Einstein condensation temperature. This dipolar shift 
depends on the angle between the 
symmetry axes of the trap and the aligned atomic dipole moments, and is 
extremal for parallel or orthogonal orientations, respectively.
The difference of both critical temperatures
exhibits most clearly the dipole-dipole interaction and
can be enhanced by increasing both the number of atoms and the anisotropy
of the trap. Applying our results to chromium atoms, which have a large magnetic dipole moment,
shows that this dipolar shift of the critical temperature could be measured in the ongoing
Stuttgart experiment.
\end{abstract}
\maketitle
\begin{figure}[t]
\begin{center}
\includegraphics[scale=0.4]{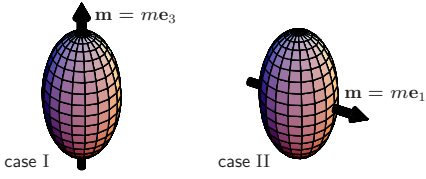}
\end{center}
\vspace*{-4mm}
\caption{\label{CASES}Symmetry axes of harmonic trap (\ref{HT}) and 
two-particle interaction (\ref{MITT}) are parallel and perpendicular 
in configuration {\rm I} and {\rm II}.}
\end{figure}

Ultracold atomic quantum gases are many-body systems in which macroscopic quantum
phenomena can be studied experimentally over a wide range of controllable interactions
\cite{Cornell,Ketterle1,Greiner1}. For the original alkali atomic 
Bose-Einstein condensates (BECs) 
it has been sufficient to describe
the dominant two-particle interaction by a local isotropic contact potential.
Recently, a new type of nonlocal anisotropic interaction has been made
accessible to detailed study by the formation of 
a BEC in a dipolar quantum gas of ${}^{52}$Cr \cite{Pfau6}. 
Chromium atoms possess a magnetic dipole moment of 6 $m_B$, where $m_B$ 
is the Bohr magneton, i.e.~it
is around six
times larger than those of alkali atoms.  It stands for a whole class of high spin atoms 
like rare earth atoms with large magnetic dipole moments 
(Dy [10~$m_B$], Ho [9~$m_B$], Eu [7~$m_B$],
Tb [10~$m_B$], Er [7~$m_B$], Mo [6~$m_B$], 
Mn [5~$m_B$], Tm [4~$m_B$], Pr [3.3~$m_B$]). 
Note that these elements have been already cooled by buffer gas and evaporative cooling 
techniques down to  mK temperatures \cite{Doyle2}.
Besides that Er [7~$m_B$] has recently been laser cooled \cite{Hanssen}.
Other many-body systems with dipolar interactions are, for instance, 
Rydberg atoms \cite{Rydberg1,Rydberg2} or atomic condensates where a strong
electric field induces electric dipole moments of the order
of $10^{-2}$ Debye \cite{Yi1}. Body-centered dipole moments in 
heteronuclear molecules are much larger with typical values
of 1 Debye, so their dipolar effects could be a few
hundred times stronger than those of chromium atoms \cite{Goral,Martikainen}.
Such a gas of ultracold heteronuclear molecules is 
produced either by sophisticated cooling and trapping 
techniques \cite{Weinstein,Meijer} 
or by photoassociation \cite{Sage,Mancini,Sengstock}.
For all those dipolar systems the
total two-particle interaction potential is modelled by
\begin{eqnarray}
\label{MITT}
&& \hspace*{-10mm}V^{({\rm int})} ( {\bf x} - {\bf x'}) = \frac{4 \pi \hbar^2 a}{M} 
\, \delta ({\bf x} - {\bf x'}) \nonumber \\
&& \hspace*{6mm}- \frac{\mu_0}{4 \pi}\, 
\left\{ \frac{3 \left[ {\bf m}\,( {\bf x}-{\bf x'})\right]^2}{|{\bf x}-{\bf x'}|^5} 
- \frac{{\bf m}^2}{|{\bf x}-{\bf x'}|^3} \right\}
\, .
\end{eqnarray} 
Here ${\bf m}$ denotes the magnetic (electric) dipole moment of the atoms,
$\mu_0$ stands for the magnetic field constant (the reciprocal electric field 
constant), and $M$ is the atomic mass. A more general 
pseudopotential for anisotropic interactions has recently 
been introduced in Refs.~\cite{Derevianko,You} which 
is  nonlocal in momentum space. Since there exists up to now no experimental evidence for
any dipolar shape resonance, where such a more general pseudopotential could be relevant,
our model (\ref{MITT}) is valid for all current and many 
future experimental situations. \\
The dimensionless measure of the strength of the dipole-dipole
interaction with respect to the s-wave scattering 
is $\epsilon_{DD} = \mu_0 m^2 M / (12 \pi \hbar^2 a)$ \cite{Pfau3}.
For the ${}^{52}$Cr condensate it has
the value $\epsilon_{DD} = 0.144$ \cite{Pfau8}, so
the magnetic dipolar interaction represents only a small 
correction to the contact interaction. Nevertheless,
it has interesting consequences due to its anisotropy, as has been
observed in a recent expansion experiment \cite{Pfau7}. 
Furthermore, the magnetic dipole-dipole interaction can be varied 
within a limited range with the help of
rotating magnetic fields as proposed by Ref.~\cite{Pfau2}. 
Combining this technique with the now known 14 Feshbach
resonances of chromium atoms \cite{Pfau8} will allow  experiments where the
interaction varies from only contact to purely dipolar. In this way, many
interesting predicted dipolar phenomena should be observed. Among them are,
for instance, the stability of the ground state of a dipolar BEC
\cite{Santos2,Eberlein1,Eberlein2} or
its excitation spectrum \cite{Santos1}.
Note that it has been recently
suggested in Refs.~\cite{Ronen1,Ronen2} that 
the $s$-wave scattering length $a$ could strongly depend on the dipole moment.
However, there it has also been shown that for
dipolar interaction strengths, which are not larger than the s-wave scattering strength, 
the latter is only rescaled
by a moderate factor and remains positive. For the calculations 
in the present work we assume that this condition is fulfilled. \\
In this letter, we investigate how the critical temperature of a dipolar BEC depends on
the dipole-dipole interaction. 
Consider an atomic gas
trapped in a cylinder-symmetric harmonic potential 
\begin{eqnarray}
\label{HT}
V({\bf x} ) = \frac{M}{2} \left[ \omega_{\perp}^2 \left( x^2 + y^2 \right) + \omega_{\parallel}^2 z^2 \right] \, ,
\end{eqnarray}
whose dipole moments ${\bf m}$ have the angle $\alpha$ with the
$z$-axis, i.e.  ${\bf m}= m \, (\sin \alpha, 0, \cos \alpha)$.
In particular, we are interested in the two extreme configurations 
{\rm I} and {\rm II} where the
symmetry axis of the dipole moments is parallel ($\alpha=0$)
and perpendicular ($\alpha=\pi/2$)
to the symmetry axis of the harmonic trap (\ref{HT}), respectively. 
Figure~\ref{CASES}
illustrates the two cases for $\omega_{\parallel} < \omega_{\perp}$, 
where the atomic dipole moments lead
to a residual attractive and repulsive interaction, respectively. At first
we show 
that the corresponding critical temperature in configuration {\rm I} and {\rm II} is shifted 
above and below the value of a pure contact interaction. Subsequently we 
suggest to determine the
difference of the critical temperatures in both configurations {\rm I} and {\rm II}
to cancel out the influence of the isotropic contact interaction.
We shall estimate which experimental parameters allow us to
enhance this signal.
For a pure contact interaction, the harmonic trap suppresses
long-wavelength fluctuations, so the leading shift of the critical
temperature can be calculated perturbatively (see, for instance,
the recent work \cite{Zobay} and the references cited therein). We expect
that this reasoning also holds for our model interaction (\ref{MITT}). 
Although the dipolar interaction is nonlocal,
its scaling properties are the same as for a
contact interaction. \\
In this letter we apply
Feynman's diagrammatic technique of many-body theory 
\cite{Mahan,Glaum} and expand the grand-canonical free energy as
\begin{eqnarray}
\raisebox{-3.5mm}{\includegraphics[scale=.9]{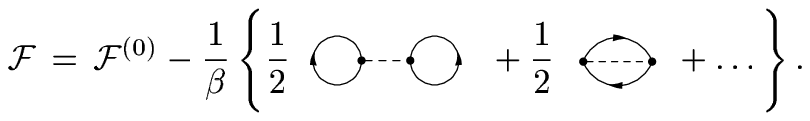}}
\label{F1}
\end{eqnarray}
The first term denotes the contribution of an ideal Bose gas 
at temperature $T = 1/ k_B \beta$ where
the harmonic trap potential (\ref{HT})  is treated semiclassically \cite{Hagen}
\begin{eqnarray}
\label{FF0}
{\cal F}^{(0)} = - \,\frac{1}{\beta (\hbar \beta \tilde{\omega})^3}\, 
\zeta_{4} \left( e^{\beta \mu} \right) \, .
\end{eqnarray}
It contains the chemical potential $\mu$,
the geometric mean of the trap frequencies 
$\tilde{\omega} = (\omega_{\parallel} \,\omega_{\perp}^2)^{1/3}$,
and the polylogarithmic function $\zeta_{a} (z) = \sum_{n=1}^{\infty} z^n/n^a $. 
The two diagrams in Eq.~(\ref{F1})
represent the direct and the exchange vacuum contribution, respectively, 
and have to be evaluated according 
to the Feynman rules:
a straight line with an arrow represents the semiclassical
interaction-free correlation function:
\begin{eqnarray}
\label{PRO}
&&\hspace*{-6mm}
\includegraphics[scale=1.0]{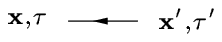}
\hspace*{2mm}\equiv  \int 
\frac{d^3 p}{(2 \pi \hbar)^3} \,
e^{i \,{\bf p} ({\bf x}-{\bf x}')/ \hbar} \, 
\nonumber \\
&&\hspace*{2mm}
\times \sum_{m=-\infty}^{\infty}
\frac{e^{- i \omega_m (\tau - \tau')}}{\beta [ - i \hbar \omega_m 
+ \frac{{\bf p}^2}{2 M} + V \left( \frac{{\bf x} + {\bf x'}}{2} \right) 
- \mu]} \, ,
\label{G}
\end{eqnarray}
where $\omega_m = 2 \pi m / \hbar \beta$ denotes the Matsubara frequency.
The two-particle interaction 
potential enters via the diagram
\begin{eqnarray}
\label{VE}
\raisebox{-1.5mm}{\includegraphics[scale=1.0]{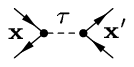}}
\hspace*{0mm}\equiv \frac{-1}{\hbar} \int_0^{\hbar \beta} d \tau \, \int d^3 x \,\int d^3 x'\,
V^{({\rm int})}({\bf x}-{\bf x'}) \, .
\end{eqnarray}
The grand-canonical free energy (\ref{F1}) is studied as a function of 
temperature $T$ for fixed particle number $N = - \partial {\cal F}/ \partial \mu$.\\
The phase transition, where a macroscopic occupation of the ground state sets in, 
occurs when the correlation function of the system diverges. From Eq.~(\ref{G})
we can see that this happens
in the interaction-free case at $\mu_c^{(0)}=0$, as the
divergence appears at the minimum of the trap potential $V({\bf x})$
for vanishing Matsubara frequency $\omega_m$ and momentum ${\bf p}$.
In the presence of a 
2-particle interaction, the full correlation function follows from a 
formula similar to (\ref{G})
where the chemical potential $\mu$ is shifted by the self-energy:
$\mu \to \mu + \hbar \,\Sigma ( {\bf p} , \omega_m ; {\bf x} )$. It
is defined by the Fourier-Matsubara transformed
\begin{eqnarray}
\label{FMT}
\int_0^{\hbar \beta} \!\!d \tau \!\int \!d^3 x' 
e^{i \omega_m \tau - i {\bf p} {\bf x'} / \hbar}  \,\Sigma \left( {\bf x} + \frac{\bf x'}{2}, \tau ; {\bf x} - \frac{\bf x'}{2} ,0 \right) 
\end{eqnarray}
of the Feynman diagrams
\begin{eqnarray}
\hspace*{-1mm}\raisebox{-4.0mm}{\includegraphics[scale=1.0]{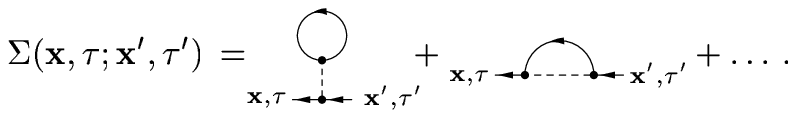}}
\end{eqnarray}
The critical chemical potential reads now \cite{NEW}
\begin{eqnarray}
\mu_c = \min_{{\bf x},{\bf p}} \left[ \frac{{\bf p}^2}{2 M} + V ({\bf x}) 
- \hbar \Sigma ( {\bf p} , 0 ; {\bf x} ) \right] \, ,
\end{eqnarray}
which leads to $\mu_c = - \hbar \, \Sigma ( {\bf 0} , 0 ; {\bf 0} )$ up to first order.
Thus, evaluating the particle number $N=N(\mu)$ at the critical chemical
potential $\mu=\mu_c$ yields the following 
first-order shift of the critical temperature 
with respect to the interaction-free critical temperature
$T_c^{(0)}= \hbar \tilde{\omega} [ N/\zeta(3)]^{1/3} /k_B$:
\begin{eqnarray}
\label{S-alfa}
\frac{\Delta T_c}{T_c^{(0)}}  =  - 
\frac{c_{\delta} a}{\lambda_c^{(0)}} + \left[ 3 \cos^2 \alpha - 1 \right] 
f \left( \frac{\omega_{\parallel}}{\omega_{\perp}} \right)\frac{\mu_0 m^2 M c_{\delta}}
{ 48 \pi \hbar^2 \lambda_c^{(0)} } \,.
\end{eqnarray} 
Here 
$\zeta(a)=\sum_{n=1}^\infty 1/n^{a}$ is the Riemann zeta function 
and $\lambda = (2 \pi \hbar^2 \beta / M)^{1/2}$
the thermodynamic de Broglie wave length.
The dimensionless prefactor $c_{\delta}$ for the $\delta$-interaction has 
the value
\begin{eqnarray}
\label{CDelta}
c_\delta = \frac{4}{3 \zeta(3)}\,\left[ \zeta \left( \frac{3}{2} \right) \,
\zeta (2) - \zeta \left( \frac{1}{2},\frac{3}{2},\frac{3}{2} \right)
\right] \approx 3.426 
\end{eqnarray}
with the generalized Riemann zeta function 
$\zeta (a,b,c) = \sum_{n=1}^\infty \sum_{n'=1}^\infty 1 / n^a {n'}^b (n+n')^c$,
whereas the dimensionless prefactor $f \left( \kappa \right) $ for 
the dipole-dipole interaction is a function of the ratio of the 
trap frequencies $\kappa=\omega_{\parallel}/\omega_{\perp}$:
%
\begin{eqnarray}
\label{FUNCY}
f ( \kappa ) = \left\{
\begin{array}{@{}ll}
{\displaystyle 
\frac{2\, \kappa^2+1}{1-\kappa^2} - 
\frac{3 \kappa^2\mbox{artanh}\, \sqrt{1-\kappa^2}}{(1-\kappa^2)^{3/2}}} & ,\, \kappa \neq 1 \\*[3.5mm]
\hspace*{1.5cm} 0 & ,\,\kappa =1  \, .
\end{array} \right. 
\end{eqnarray}
This function was already used in Ref.~\cite{Pfau3} to describe the mean-field
magnetic dipole-dipole energy for a cylindrically symmetric BEC. 
Note that $f(\kappa)$ tends asymptotically to $-2$ for $\kappa \to \infty$
and 1 for $\kappa \to 0$, respectively.\\
\begin{figure}[t]
\begin{center}
\epsfig{file=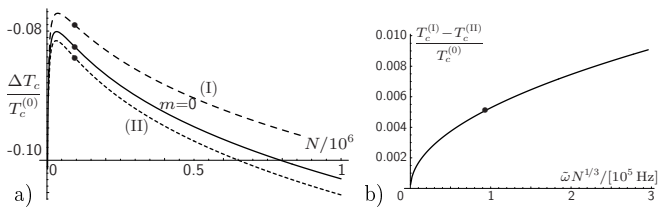,width=\columnwidth} 
\end{center}
\vspace*{-4mm}
\caption{\label{RES}a) Shift of the critical temperature $\Delta T_c$ with 
respect to the interaction-free critical temperature $T_c^{(0)}$ for a 
${}^{52}$Cr gas in a harmonic trap with frequencies 
$\omega_{\parallel} = 2 \pi \cdot 138\,\mbox{Hz}$ and 
$\omega_{\perp} = 2 \pi \cdot 486\,\mbox{Hz}$: without (straight line) and 
with (dashed lines) magnetic
dipole-dipole interaction for the configurations {\rm I} and {\rm II} of Figure~\ref{CASES}.
b) Difference of the temperature shifts increases with particle number $N$
and geometric mean frequency $\tilde{\omega}$. 
The respective dots
indicate the parameters of the Stuttgart ${}^{52}$Cr experiment.}
\end{figure}
The general physical implications of our first-order perturbative result (\ref{S-alfa})
are as follows. Without the dipole-dipole interaction, i.e. at
$m=0$, 
the critical temperature is shifted downwards with the dimensionless prefactor 
(\ref{CDelta}). This result for the 
isotropic contact interaction was originally
derived in Ref.~\cite{Giorgini} within a mean-field approach and
confirmed experimentally by investigating the onset of Bose-Einstein condensation in the 
hyperfine ground state of ${}^{87}$Rb \cite{Aspect}. 
Our result (\ref{S-alfa}) for $m \neq 0$ shows that
the critical temperature is increased in configuration {\rm I} ($\alpha=0$) 
twice as much as decreased
in configuration {\rm II} ($\alpha = \pi/2$) of Figure~\ref{CASES} for the 
trap anisotropy $\omega_{\parallel} < \omega_{\perp}$,
due to the dipole-dipole interaction. 
Of course, changing the angle $\alpha$ allows us
to tune the dipolar effect between these maximal and minimal values. 
In particular, we read off from Eq.~(\ref{S-alfa})
that the dipolar shift of the critical temperature vanishes for the magic angle
$\alpha_0 = \arccos \,( 1/  \sqrt{3} ) = 54.7^{\,\mbox{o}}$ \cite{Pfau2}.\\
Now we discuss the consequences of our results
for the ongoing experiments on the Bose-Einstein
condensation of ${}^{52}$Cr-atoms at the University of Stuttgart, where the 
trap frequencies are $\omega_1 = 2 \pi \cdot 581\,\mbox{Hz}$, 
$\omega_2 = 2 \pi \cdot 406\,\mbox{Hz}$, 
$\omega_3 = 2 \pi \cdot 138\,\mbox{Hz}$, so that the geometric
mean frequency is $\tilde{\omega}=2\pi \cdot 319\,\mbox{Hz}$.
The total number of atoms is $N=100\,000$, yielding
an interaction-free critical temperature
of about $T_c^{(0)} = 670 \, \mbox{nK}$. The corresponding finite-size 
correction was calculated in Refs. \cite{Grossmann,Dalfovo}
\begin{eqnarray}
\label{FS}
\left( \frac{\Delta T_c}{T_c^{(0)}}\right)_{FS} = \,
- \frac{\zeta(2) \overline{\omega}}{2 \zeta^{2/3}(3) \tilde{\omega} N^{1/3}} \, ,
\end{eqnarray}
where $\overline{\omega}$ is the arithmetic mean of the trap frequencies 
$\overline{\omega} \approx
(\omega_1+\omega_2+\omega_3)/3 = 2 \pi \cdot 375\,\mbox{Hz}$. Thus, the 
finite-size correction of $T_c^{(0)}$ in the Stuttgart experiment amounts 
to $-1.8$ \%. This is to be compared with a shift of the critical temperature
due to the contact and the magnetic dipole-dipole 
interaction following from formula (\ref{S-alfa}).\\
\begin{figure}[t]
\begin{center}
\epsfig{file=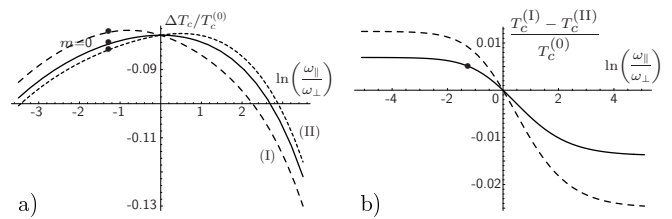,width=\columnwidth} 
\end{center}
\vspace*{-4mm}
\caption{\label{RES-kappa}a) Shift of the critical temperature $\Delta T_c$ 
with respect to the interaction-free critical temperature $T_c^{(0)}$ for 
$N=10^5$ ${}^{52}$Cr-atoms gas versus anisotropy parameter 
$\omega_{\parallel}/\omega_{\perp}$ of the harmonic 
trap without (straight line) and with (dashed lines) magnetic
dipole-dipole interaction for the  configurations {\rm I} and {\rm II} of Figure~\ref{CASES}.
b) Difference of the temperature shifts versus anisotropy parameter 
$\omega_{\parallel}/\omega_{\perp}$ for
$\tilde{\omega} N^{1/3} = 0.93 \cdot 10^5 \,\mbox{Hz}$ (solid line) 
and $\tilde{\omega} N^{1/3} = 3 \cdot 10^5 \,\mbox{Hz}$ 
(dashed line). The respective dots
indicate the parameters of the Stuttgart ${}^{52}$Cr experiment.}
\end{figure}
The s-wave scattering length of the ${}^{52}$Cr-atoms is 
$a = 105\,a_B$  \cite{Pfau8}, i.e. roughly 2 \% of the 
thermodynamic de Broglie wave length  
$\lambda_c^{(0)}= 5\,598\,a_B$. The corresponding 
downwards shift of the critical temperature amounts then to
$6.4$ \%. This is modified by the 
dipole-dipole interaction depending on the experimental set-up. 
In the Stuttgart experiment, the trap potential has actually
three different frequencies but, since $\omega_1$ and $\omega_2$ are not far apart, we may 
identify  $\omega_{\perp} = \sqrt{\omega_1 \omega_2}
= 2 \pi \cdot 486\,\mbox{Hz}$ and
$\omega_{\parallel} = \omega_3$, yielding
$\omega_{\parallel} / \omega_{\perp} =0.284$.
In configuration {\rm I} we obtain from (\ref{FUNCY}) 
the result $f ( \omega_{\parallel} / \omega_{\perp} )=0.733$, which
leads via Eq.~(\ref{S-alfa}) to an extra upward shift of the critical 
temperature  by $0.34$ \% due to the magnetic dipole-dipole interaction. 
The downward shift in configuration {\rm II} is half as big. Figure \ref{RES}a plots
the resulting total shift of the
critical temperature $\Delta T_c$ for the ${}^{52}$Cr gas with respect to the
interaction-free critical temperature $T_c^{(0)}$ versus the particle number $N$.
Both the finite-size corrections and the contact interaction lead to the main shift, 
on top of which the small dipolar effect is seen. Figure \ref{RES-kappa}a
shows how the same shifts depend for $N=10^5$ chromium
atoms on the anisotropy parameter $\omega_{\parallel}/\omega_{\perp}$.
The directions of the shifts change sign at the isotropy point $\omega_{\parallel} =\omega_{\perp}$.  \\
The above results suggest to plot the difference between the critical temperatures of the two  
configurations {\rm I} and {\rm II}. This eliminates the isotropic effects of both
the contact interaction and the finite-size correction, thus it
exhibits most clearly the magnetic dipole-dipole interaction.
For $N=10^5$ and $\tilde{\omega}=2\pi \cdot 319\,\mbox{Hz}$, 
the difference amounts to a
net effect of 0.51 \% of the interaction-free critical temperature $T_c^{(0)}$.
One possibility to enhance the difference of the critical
temperatures is to choose a convenient anisotropy strength 
$\omega_{\parallel}/\omega_{\perp}$ of the harmonic trap potential
as seen in Figure \ref{RES-kappa}b. Furthermore,
this dipolar effect increases with the total atom number 
$N$ and the geometric mean frequency $\tilde{\omega}$
as shown in Figure~\ref{RES}b, which therefore needs to be as large as possible.
An experimentally feasible increase of the particle density in the trap
could lead to a dipolar effect of more than 2 \%, see dashed curve of Figure 3b. \\
At present, the best experiments which measure the critical temperature of a 
Bose-Einstein condensate involve
error bars of 5 \% which represent the total systematic and statistical 
errors \cite{Pfau6,Aspect}. 
The systematic errors, however, can be eliminated
by our suggestion to measure the difference of two critical temperatures. 
We expect that the remaining statistical errors can be reduced
to 1 \%
by averaging both atom number and critical temperature over many measurements. \\
So far, the dipolar nature of the chromium BEC has only been resolved in expansion
experiments \cite{Pfau7}. The analysis of this letter shows that it could be
possible to detect a signal of the underlying magnetic dipole-dipole interaction
also by measuring the critical temperature.
Furthermore, our results will be useful for other dipolar systems with a tunable
dipole moment, like heteronuclear molecules in low vibrational states 
\cite{Weinstein,Meijer,Sage,Mancini,Sengstock}, where the dipolar effect will
be larger.\\
We thank A. Berra and S. Kling for discussions and 
DFG Priority Program SPP 1116 as well as SFB/TR 21.
\end{document}